\documentclass[aps,twocolumn,showpacs,preprintnumbers,amsmath,amssymb,superscriptaddress,floatfix,nofootinbib]{revtex4}

\usepackage{graphicx}
\usepackage{amsmath}
\usepackage{amsfonts}
\usepackage{amssymb}
\usepackage{color}

\newcommand{\Slash}[1]{\ooalign{\hfil/\hfil\crcr$#1$}}

\begin{document}
\title{Nucleon resonances in the $\gamma p \to \phi K^+ \Lambda$ reaction near threshold}

\author{Qi-Fang L\"{u}} \affiliation{Department of Physics, Zhengzhou University, Zhengzhou, Henan 450001, China}

\author{Rong Wang}
\affiliation{Institute of Modern Physics, Chinese Academy of
Sciences, Lanzhou 730000, China} \affiliation{University of Chinese
Academy of Sciences, Beijing 100049, China}

\author{Ju-Jun Xie}
\affiliation{Institute of Modern Physics, Chinese Academy of
Sciences, Lanzhou 730000, China} \affiliation{University of Chinese
Academy of Sciences, Beijing 100049, China} \affiliation{Research
Center for Hadron and CSR Physics, Institute of Modern Physics of
CAS and Lanzhou University, Lanzhou 730000, China}

\affiliation{State Key Laboratory of Theoretical Physics, Institute
of Theoretical Physics, Chinese Academy of Sciences, Beijing 100190,
China}

\author{Xu-Rong Chen}
\affiliation{Institute of Modern Physics, Chinese Academy of
Sciences, Lanzhou 730000, China} \affiliation{University of Chinese
Academy of Sciences, Beijing 100049, China}

\author{De-Min Li} \email{lidm@zzu.edu.cn}
\affiliation{Department of Physics, Zhengzhou University, Zhengzhou, Henan 450001, China}

\begin{abstract}

We investigate the $\gamma p \to \phi K^+ \Lambda$ reaction near
threshold within an effective Lagrangian approach and the isobar
model. Various nucleon resonances caused by the $\pi$ and $\eta$
meson exchanges and background contributions are considered. It is
shown that the contribution from the $N^*(1535)$ resonance caused by
the $\eta$ meson exchange plays the predominant role. Hence, this
reaction provides a good new platform to study the $N^*(1535)$
resonance. The predicted total cross section and specific features
of angular distributions can be tested by future experiments.

\end{abstract}
\date{\today}

 \maketitle

\section{Introduction}{\label{introduction}}

The study of nucleon resonances in meson production reaction is an
interesting topic in light hadron physics. Among these resonances,
the negative parity nucleon excited state, $N^*(1535)$ (spin-parity
$J^P = 1/2^-$), has been a controversial resonance for many years.
In the traditional three-quark constituent model, it should be the
lowest spatially excited nucleon state with one quark in a
$p-$wave~\cite{pdg2014,Capstick:1986bm}. However, the $N^*(1440)$
($J^P =1/2^+$) has in fact a much lower mass, despite requiring two
units of excitation energy. This is the long-standing mass inversion
problem of the nucleon resonance spectrum.

Furthermore, the $N^*(1535)$ resonance couples strongly to those
strangeness channels~\cite{Xie:2010md}. Besides the $\eta N$
channel~\cite{pdg2014}, it also couples strongly to $\eta^{\prime}
N$~\cite{Cao:2008st,Zhong:2011ht,Huang:2012xj}, $\phi
N$~\cite{Xie:2007qt,Dai:2011yr}, and $K
\Lambda$~\cite{Liu:2005pm,Liu:2006ym} channels. Moreover, it is
found that the $N^*(1535)$ state is dynamically generated within a
chiral unitary coupled channel approach, with its mass, width, and
branching ratios in fair agreement with the experimental
results~\cite{osetprc65,kaisernpa612,nievesprd64,garciaplb582}. This
approach shows that the couplings of the $N^*(1535)$ resonance to
the $K \Sigma$, $\eta N$, and $K \Lambda$ channels could be large
compared to that for the $\pi N$ channel.

The mass inversion problem could be understood if there were
significant five-quark ($uud s\bar{s}$) components in the wave
function of the $N^*(1535)$ resonance~\cite{Zou:2009zz,Zou:2010tc},
which would also provide a natural explanation of the large
couplings of the the $N^*(1535)$ resonance to the strangeness
$K\Lambda$, $K\Sigma$, $N\eta'$, and $N\phi$ channels. It would
furthermore lead to an improvement in the description of the
$N^*(1535)$ helicity amplitudes~\cite{An:2008xk,An:2009zza}. In this
paper, we wish to argue that the $N^*(1535)$ resonance might play an
important role in the associated strangeness production of the
$\gamma p \to \phi K^+ \Lambda$ reaction. Since this reaction needs
to create two $s\bar{s}$ quark pairs from the vacuum, its total
cross sections should be small, which is why not much attention has
been paid to it on either the theoretical or experimental sides.
However, because of the week interactions of $\phi K^+$ and $\phi
\Lambda$, the $\gamma p \to \phi K^+ \Lambda$ reaction near
threshold provides a good new platform to study the $N^*(1535)$
resonance decaying to $K^+ \Lambda$.

The couplings of the $N^*(1535)$ resonance to the $K\Lambda$ and
$\eta N$ channels and the ratio of $g_{K\Lambda N^*(1535)}$ to
$g_{\eta N N^*(1535)}$, $R \equiv | g_{K \Lambda N^*(1535)}/g_{\eta
N N^*(1535)} |$, have been intensively studied within various
theoretical approaches. By analyzing the $J/\psi \to \bar{p}K^+
\Lambda$ and $J/\psi \to \bar{p}\eta p$ experimental data,
Ref.~\cite{Liu:2005pm} gives  $R = 1.3 \pm 0.3$. From the latest and
largest photoproduction database by using the isobar model,
Ref.~\cite{Mart:2013ida} gives $R = 0.460 \pm 0.172$.  From the
$J/\psi$ decays within the chiral unitary approach,
Ref.~\cite{Geng:2008cv} gives $R= 0.5 \sim 0.7$. Based on the
partial wave analysis of kaon photonproduction,
Ref.~\cite{Sarantsev:2005tg} gives $R= 0.42 \sim 0.73$. The result
of the $s$-wave $\pi N$ scattering analysis within a unitarized
chiral effective Lagrangian indicates $|g_{ g_{K \Lambda
N^*(1535)}}|^2 > |g_{\eta N N^*(1535)}|^2$~\cite{Bruns:2010sv}. The
coupled-channels calculation predicts $R = 0.8 \sim
2.6$~\cite{Penner:2002ma}. Study on the partial decay widths of the
$N^*(1535)$ resonance to the pseudoscalar mesons and octet baryons
within a chiral constituent quark model shows $R = 0.85 \pm
0.06$~\cite{An:2011sb}. In a very recent analysis of the $\pi^- p
\to K^0 \Lambda$ reaction, a value $R = 0.71 \pm 0.10$ is
obtained~\cite{Wu:2014yca}. Obviously, theoretical predictions on
$R$ are not completely consistent with each other, thus, it is still
worth studying the coupling constant $g_{K\Lambda N^*(1535)}$ in
different ways.

In the present work, we investigate the role of nucleon resonances
in the $\gamma p \to \phi K^+ \Lambda$ reaction near threshold in
the framework of an effective Lagrangian approach and the isobar
model. Initial interaction between incoming photons and protons is
modeled by an effective Lagrangian which is based on the exchange of
the $\pi$, $\eta$, and kaon mesons. The $K^+\Lambda$ production
proceeds via excitation of the $N^*(1535)$, $N^*(1650)$,
$N^*(1710)$, and $N^*(1720)$ intermediate nucleon resonances which
have appreciable branching ratios for the decay into the $K^+
\Lambda$ channel.

This article is organized as follows. In the next section, we will
present the formalism and ingredients for present calculations, then
numerical results and discussions are given in Sect. III. A short
summary is given in the last section.

\section{Formalism and ingredients}{\label{formalism}}

\begin{figure*}[htbp]
\begin{center}
\includegraphics[scale=0.5]{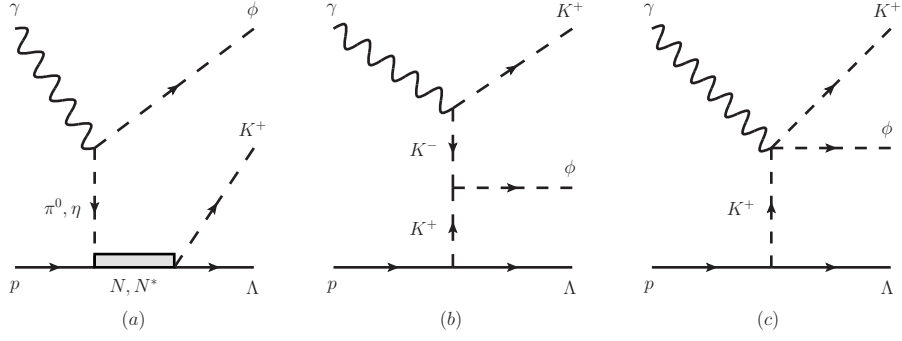} \caption{Feynman
diagrams for the $\gamma p \to \phi K^+ \Lambda$ reaction.}
\label{diagram}
\end{center}
\end{figure*}

We study the $\gamma p \to \phi K^+ \Lambda$ reaction near threshold
within an effective Lagrangian approach and the isobar model, which
has been extensively applied to the study of scattering
processes~\cite{Tsushima:1998jz,Sibirtsev:2005mv,Liu:2006tf,Maxwell:2012zz,Liu:2012kh,Xie:2013wfa,Liu:2011sw,Liu:2012ge,Lu:2013jva,Lu:2014rla,Xie:2013mua,Xie:2010yk}.
The basic tree level Feynman diagrams for the $\gamma p \to \phi K^+
\Lambda$ reaction are depicted in Fig.~\ref{diagram}. It is assumed
that the $K \Lambda$ final state is produced by the decay of the
intermediate nucleon resonances as the result of the $\pi$ and
$\eta$ meson exchange [Fig.~\ref{diagram} (a)]. Moreover, the
background contributions including the $s$-channel nucleon pole,
$t$-channel $K$ exchange [Fig.~\ref{diagram} (b)], and contact term
[Fig.~\ref{diagram} (c)] are also considered.

To compute the amplitudes of these diagrams shown in
Fig.~\ref{diagram}, the effective Lagrangian densities for relevant
interaction vertexes are needed. We use the commonly employed
Lagrangian densities for $\pi N N$, $\eta N N$, and $K \Lambda N$ as
follows~\cite{Liang:2004sd}:

\begin{eqnarray}
{\cal L}_{\pi N N}  &=& - \frac{g_{\pi N N}}{2m_N} \bar{N} \gamma_5
\gamma_{\mu} \vec\tau \cdot \partial^{\mu} \vec\pi N, \label{pin} \\
{\cal L}_{\eta N N} &=& - \frac{g_{\eta N N}}{2m_N} \bar{N} \gamma_5
\gamma_{\mu} \partial^{\mu} \eta N, \label{etan} \\
{\cal L}_{K \Lambda N} &=& - \frac{g_{K \Lambda N}}{m_N+m_{\Lambda}}
\bar{N} \gamma_5 \gamma_{\mu}
\partial^{\mu}K \Lambda + {\rm h.c.}. \label{kn}
\end{eqnarray}

The coupling constants in the above Lagrangian densities are taken
as~\cite{Xie:2007qt,Xie:2013wfa,Lu:2014rla}: $g_{\pi NN} = 13.45$, $g_{K N \Lambda} =
-13.98$, and $g_{\eta N N} = 2.24$.

For the $\phi$ meson and photon couplings, we use the interaction
Lagrangian densities as used in Refs.~\cite{Xie:2010yk,Ryu:2012tw},
\begin{eqnarray}
{\cal L}_{\phi \gamma \pi}  &=& \frac{e}{m_{\phi}} g_{\phi \gamma
\pi} \epsilon^{\mu \nu \alpha \beta } \partial_{\mu} \phi_{\nu}
\partial_{\alpha} A_{\beta} \pi, \label{phipi} \\
{\cal L}_{\phi \gamma \eta}  &=& \frac{e}{m_{\phi}} g_{\phi \gamma
\eta} \epsilon^{\mu \nu \alpha \beta } \partial_{\mu} \phi_{\nu}
\partial_{\alpha} A_{\beta} \eta, \label{phieta} \\
{\cal L}_{\gamma K K} &=& -i e (\partial^{\mu}K^-K^+ -
\partial^{\mu}K^+K^-)A_{\mu}, \label{kk} \\
{\cal L}_{\phi K K} &=& -i g_{\phi K K} (\partial^{\mu}K^-K^+ -
\partial^{\mu}K^+K^-){\phi}_{\mu} \label{phikk}
\end{eqnarray}
where $e = \sqrt{4\pi \alpha}$ ($\alpha = 1/137.036$ is the
fine-structure constant), and $A_{\mu}$ is the photon field.

We also need the interaction Lagrangian densities involving nucleon
resonances ($\equiv R$)~\cite{Zou:2002yy},
\begin{eqnarray}
{\cal L}_{\pi N R} &=& i g_{\pi N R} \bar{R} \vec\tau \cdot \vec\pi
N + {\rm h.c.}, \label{pin1535} \\
{\cal L}_{\eta N R} &=& i g_{\eta N R} \bar{R} \eta N + {\rm h.c.},
\label{etan1535} \\
{\cal L}_{K \Lambda R} &=& i g_{K \Lambda R} \bar{R} K \Lambda +
{\rm h.c.}, \label{kn1535} \label{etan1535}
\end{eqnarray}
for $J^p(\frac{1}{2}^-)$ nucleon resonances $N^*(1535)$ and
$N^*(1650)$, while
\begin{eqnarray}
{\cal L}_{\pi N R} &=& - \frac{g_{\pi N R}}{m_N+m_R} \bar{R}
\gamma_5 \gamma_{\mu} \vec\tau \cdot \partial^{\mu} \vec\pi N + {\rm
h.c.}, \label{pin1710} \\
{\cal L}_{\eta N R} &=& - \frac{g_{\eta N R}}{m_N+m_R} \bar{R}
\gamma_5 \gamma_{\mu} \partial^{\mu} \eta N+ {\rm h.c.},
\label{etan1710} \\
{\cal L}_{K \Lambda R} &=& - \frac{g_{K \Lambda R}}{m_{\Lambda}+m_R}
\bar{R} \gamma_5 \gamma_{\mu}
\partial^{\mu}K \Lambda + {\rm h.c.}, \label{kn1710}
\end{eqnarray}
for $J^p(\frac{1}{2}^+)$ nucleon resonance $N^*(1710)$, and
\begin{eqnarray}
{\cal L}_{\pi N R} &=& - \frac{g_{\pi N R}}{m_{\pi}} \bar{R}_{\mu}
\vec\tau \cdot \partial^{\mu} \vec\pi N + {\rm h.c.},
\label{pin1720} \\
{\cal L}_{\eta N R} &=& - \frac{g_{\eta N R}}{m_{\eta}}
\bar{R}_{\mu}
\partial^{\mu} \eta N + {\rm h.c.}, \label{etan1720} \\
{\cal L}_{K \Lambda R} &=& - \frac{g_{K \Lambda R}}{m_K}
\bar{R}_{\mu}
\partial^{\mu} K \Lambda + {\rm h.c.} , \label{kn1720}
\end{eqnarray}
for $J^p(\frac{3}{2}^+)$ nucleon resonance $N^*(1720)$.

The coupling constants in the above Lagrangian densities can be
determined from the partial decay widths,
\begin{eqnarray}
\Gamma[\phi \to \pi \gamma] &=& \frac{e^2g^2_{\phi \gamma \pi}}{12\pi} \frac{|\vec{p}_{\pi \gamma}|^3}{m_{\phi}^2} , \\
\Gamma[\phi \to \eta \gamma] &=& \frac{e^2g^2_{\phi \gamma \eta}}{12\pi} \frac{|\vec{p}_{\eta \gamma}|^3}{m_{\phi}^2} , \\
\Gamma[\phi \to K^+ K^-] &=& \frac{g^2_{\phi K K}}{6\pi} \frac{|\vec{p}_{K^+ K^-}|^3}{m_{\phi}^2} ,
\end{eqnarray}
for the $\phi$ meson,
\begin{eqnarray}
\Gamma[R \to N \pi] &=& \frac{3g^2_{\pi N R}}{4\pi} \frac{(E_N+m_N)}{m_R} |\vec{p}_{N\pi}|, \\
\Gamma[R \to N \eta] &=& \frac{g^2_{\eta N R}}{4\pi} \frac{(E_N+m_N)}{m_R} |\vec{p}_{N\eta}|, \\
\Gamma[R \to \Lambda K] &=& \frac{g^2_{K \Lambda R}}{4\pi} \frac{(E_{\Lambda}+m_{\Lambda})}{m_R} |\vec{p}_{\Lambda K}|,
\end{eqnarray}
for $J^p(\frac{1}{2}^-)$ nucleon resonances $N^*(1535)$ and
$N^*(1650)$,
\begin{eqnarray}
\Gamma[R \to N \pi] &=& \frac{3g^2_{\pi N R}}{4\pi} \frac{(E_N-m_N)}{m_R(m_R+m_N)} |\vec{p}_{N\pi}|, \\
\Gamma[R \to N \eta] &=& \frac{g^2_{\eta N R}}{4\pi} \frac{(E_N-m_N)}{m_R(m_R+m_N)} |\vec{p}_{N\eta}|, \\
\Gamma[R \to \Lambda K] &=& \frac{g^2_{K \Lambda R}}{4\pi} \frac{(E_{\Lambda}-m_{\Lambda})}{m_R(m_R+m_{\Lambda})} |\vec{p}_{\Lambda K}|,
\end{eqnarray}
for $J^p(\frac{1}{2}^+)$ nucleon resonance $N^*(1710)$, and
\begin{eqnarray}
\Gamma[R \to N \pi] &=& \frac{g^2_{\pi N R}}{4\pi} \frac{(E_N+m_N)}{m_R m_{\pi}^2} |\vec{p}_{N\pi}|^3, \\
\Gamma[R \to N \eta] &=& \frac{g^2_{\eta N R}}{12\pi} \frac{(E_N+m_N)}{m_R m_{\eta}^2} |\vec{p}_{N\eta}|^3, \\
\Gamma[R \to \Lambda K] &=& \frac{g^2_{K \Lambda R}}{12\pi} \frac{(E_{\Lambda}+m_{\Lambda})}{m_R m_K^2} |\vec{p}_{\Lambda K}|^3,
\end{eqnarray}
for $J^p(\frac{3}{2}^+)$ nucleon resonance $N^*(1720)$, where
\begin{eqnarray}
|\vec{p}_{f_1 f_2}| &=& \frac{\lambda^{\frac{1}{2}}(m_i^2,m_{f_1}^2,m_{f_2}^2)}{2m_i},
\end{eqnarray}
where $m_i$ denotes the mass of the initial state, $m_{f_1}$ and
$m_{f_2}$ are the masses of two final states, and $\lambda$ is the
K\"{a}llen function with $\lambda(x,y,z) = (x-y-z)^2 - 4yz$. The
obtained results for the coupling constants are listed in
Tab.~\ref{tab1}, while the coupling constant $g_{K \Lambda
N^*(1535)}$ will be discussed below.

\begin{table}[htbp]
\begin{center}
\caption{ \label{tab1} Relevant parameters used in the present
calculation. The widths and branching ratios are taken from Particle
Data Group~\cite{pdg2014}.} \footnotesize
\begin{tabular}{ccccc}
\hline
\hline
State &  Width & Decay     & Adopted          & $g^2/4\pi$    \\
      &  (MeV) & channel   & branching ratio  &               \\
\hline
$\phi $    &  4.26       & $\pi \gamma$    & $1.27 \times 10^{-3}$  & $1.56 \times 10^{-3}$ \\
           &             & $\eta \gamma$   & $1.31 \times 10^{-2}$  & $4.01 \times 10^{-2}$\\
           &             & $K^+ K^- $      & 0.49                   & $1.59$ \\
$N^*(1535)$&  150        & $N \pi$         & 0.45                   & $3.68 \times 10^{-2}$\\
           &             & $N \eta$        & 0.42                   & $0.28$\\
$N^*(1650)$&  150        & $N \pi$         & 0.70                   & $5.22 \times 10^{-2}$\\
           &             & $N \eta$        & 0.10                   & $3.57 \times 10^{-2}$\\
           &             & $\Lambda K$     & 0.07                   & $4.36 \times 10^{-2}$\\
$N^*(1710)$&  100        & $N \pi$         & 0.13                   & $7.18 \times 10^{-2}$\\
           &             & $N \eta$        & 0.20                   & $0.97$\\
           &             & $\Lambda K$     & 0.15                   & $2.98$\\
$N^*(1720)$&  250        & $N \pi$         & 0.11                   & $2.04 \times 10^{-3}$\\
           &             & $N \eta$        & 0.04                   & $0.11$\\
           &             & $\Lambda K$     & 0.08                   & $0.49$\\
\hline
\hline
\end{tabular}
\end{center}
\end{table}

Since the hadrons are not point-like particles, the form factors are
also needed. We adopt the dipole form factor for exchanged
mesons~\cite{Xie:2007qt,Oh:2007jd,Gao:2010hy},
\begin{equation}
F_M(q^{2}_{ex},M_{ex}) = (\frac{\Lambda_M^2 - M^2_{ex}}{\Lambda_M^2
- q^2_{ex}})^2.
\end{equation}

The form factor for the exchanged baryons is taken
as~\cite{Feuster:1997pq,Shklyar:2005xg},
\begin{equation}
F_B(q^{2}_{ex},M_{ex}) = \frac{\Lambda_B^4}{\Lambda_B^4 +
(q^{2}_{ex}-M^2_{ex})^2},
\end{equation}
where the $q_{ex}$ and $M_{ex}$ are the four-momentum and the mass
of the exchanged hadron, respectively. In our present calculation,
we use the cut off parameters $\Lambda_{\pi} = \Lambda_{\eta} = 1.3$
GeV for $\pi$ and $\eta$ mesons~\cite{Xie:2007qt}, $\Lambda_K = 0.8$
GeV for $K$ meson~\cite{Oh:2007jd,Gao:2010hy}, and $\Lambda_N =
\Lambda_{N^*(1535)} = \Lambda_{N^*(1650)} = \Lambda_{N^*(1710)}=
\Lambda_{N^*(1720)} = 2.0$ GeV for baryons~\cite{Xie:2007qt}.

The propagators for exchanged $\pi$, $\eta$ and $K$ meson used in
our calculation is,
\begin{equation}
G_{\pi, \eta, K}(q)=\frac{i}{q^2-m_{\pi, \eta, K}^2}.
\end{equation}
For the propagator of spin-1/2 baryon, we use
\begin{equation}
G_{\frac{1}{2}}(q) = \frac{i(\Slash q + M)}{q^2 - M^2+ iM \Gamma}.
\end{equation}
For the propagator of spin-3/2 baryon, it can be taken as,
\begin{equation}
G_{\frac{3}{2}}^{\mu\nu}(q) =  \frac{i(\Slash q + M)
P^{\mu\nu}(q)}{q^2 - M^2+ iM\Gamma},
\end{equation}
with
\begin{eqnarray}
P^{\mu\nu}(q) &=& -g^{\mu\nu} + \frac{1}{3}\gamma^{\mu}\gamma^{\nu} + \frac{1}{3M}(\gamma^{\mu}q^{\nu}-\gamma^{\nu}q^{\mu}) \nonumber  \\
&& + \frac{2}{3M^2}q^{\mu}q^{\nu},
\end{eqnarray}
where $q$, $M$, and $\Gamma$ stand for the four-momentum, mass, and
total width of the intermediate nucleon resonance, respectively.

From the above effective Lagrangian densities, the scattering
amplitudes for the $\gamma p \to \phi K^+ \Lambda$ reaction can be
obtained straightforwardly. For nucleon pole due to the $\pi$
exchange can be written as,
\begin{eqnarray}
{\cal M}_{N} & = & \frac{i g_{K \Lambda N} g_{\pi N N} g_{\phi
\gamma \pi}F_{\pi}(q_{\pi}^2,m_{\pi}) F_N(q_N^2,m_N)}{2m_N
(m_N+m_{\Lambda})
m_{\phi}} \nonumber \\
&&  \times \bar{u} (p_5,s_5) \gamma_5 \Slash{p}_4
G_{1/2}(q_N) \gamma_5 \Slash {q}_{\pi} u(p_2,s_2)G_{\pi}(q_{\pi})  \nonumber \\
&&  \times \epsilon^{\mu \nu \alpha \beta} p_{3\mu}
\varepsilon^*_{\nu}(p_3,s_3) p_{1 \alpha}
\varepsilon_{\beta}(p_1,s_1),
\end{eqnarray}

For $J^p(\frac{1}{2}^-)$ nucleon resonances $N^*(1535)$ and
$N^*(1650)$,
\begin{eqnarray}
{\cal M}_R & = &  \frac{-i g_{K \Lambda R} g_{\pi N R} g_{\phi
\gamma \pi} F_{\pi}(q_{\pi}^2,m_{\pi}) F_R(q_R^2,m_R)}{m_{\phi}} \nonumber \\
&&  \times \bar{u} (p_5,s_5) G_{1/2}(q_R) u(p_2,s_2) G_{\pi}(q_{\pi})  \nonumber \\
&&  \times \epsilon^{\mu \nu \alpha \beta} p_{3\mu}
\varepsilon^*_{\nu}(p_3,s_3) p_{1\alpha}
\varepsilon_{\beta}(p_1,s_1),
\end{eqnarray}

For $J^p(\frac{1}{2}^+)$ nucleon resonance $N^*(1710)$,
\begin{eqnarray}
{\cal M}_R & = & \frac{i g_{K \Lambda R} g_{\pi N R} g_{\phi \gamma
\pi} F_{\pi}(q_{\pi}^2,m_{\pi})F_R(q_R^2,m_R) }{(m_N+m_R)
(m_{\Lambda}+m_R)
m_{\phi}} \nonumber \\
&& \times  \bar{u} (p_5,s_5) \gamma_5 \Slash {p}_4
G_{1/2}(q_R) \gamma_5 \Slash {q}_{\pi} u(p_2,s_2) G_{\pi}(q_{\pi}) \nonumber \\
&& \times \epsilon^{\mu \nu \alpha \beta} p_{3\mu}
\varepsilon^*_{\nu}(p_3,s_3) p_{1\alpha}
\varepsilon_{\beta}(p_1,s_1),
\end{eqnarray}

For $J^p(\frac{3}{2}^+)$ nucleon resonance $N^*(1720)$,
\begin{eqnarray}
{\cal M}_R & = & \frac{i g_{K \Lambda R} g_{\pi N R} g_{\phi \gamma
\pi} F_{\pi}(q_{\pi}^2,m_{\pi})  F_R(q_R^2,m_R)}{m_{\pi} m_K
m_{\phi}} \nonumber\\ &&  \times \bar{u} (p_5,s_5) p_{4\mu}
G_{3/2}^{\mu\nu}(q_R) q_{\pi \nu} u(p_2,s_2) G_{\pi}(q_{\pi})
\nonumber\\ &&  \times \epsilon^{\mu \nu \alpha \beta} p_{3\mu}
\varepsilon^*_{\nu}(p_3,s_3) p_{1\alpha}
\varepsilon_{\beta}(p_1,s_1),
\end{eqnarray}
where $p_i~(i = 1,2,3,4,5)$ stand the four-momenta for photon,
proton, $\phi$, $K^+$, and $\Lambda$, respectively, and $s_i~(i
=1,2,3,5)$ represent the spin projection of photon, proton, $\phi$,
and $\Lambda$, respectively. $q_{\pi} = p_1-p_3$ is the
four-momentum for exchanged $\pi$ meson, and $q_R = p_4+p_5$ is the
four-momentum for exchanged nucleon resonance. The amplitudes due to
the $\eta$ exchange are similar, and can be obtained by changing
$\pi$ to $\eta$ from the above equations.

We also give the amplitudes of the $K$ exchanged Feynman diagram
shown in Fig.~\ref{diagram} (b) as follows,
\begin{eqnarray}
{\cal M}_K & = & \frac{e g_{K \Lambda N} g_{\phi K
K}}{m_N+m_{\Lambda}} F_K(q_{K^-}^2,m_{K^-}) F_K(q_{K^+}^2,m_{K^+})
\nonumber \\
&& \times \bar{u} (p_5,s_5) \gamma_5 \Slash {q}_{K^+} u(p_2,s_2)
G_{K}(q_{K^+}) \nonumber \\
&& \times (q_{K^-}-q_{K^+}) \cdot \varepsilon^*(p_3,s_3)
G_{K}(q_{K^-}) \nonumber \\
&& \times (p_4-q_{K^-}) \cdot \varepsilon(p_1,s_1),
\end{eqnarray}
$q_{K^-} = p_1-p_4$ is the four-momentum for exchanged $K^-$ meson, and $q_{K^+} =
p_2-p_5$ is the four-momentum for exchanged $K^+$ meson.

The contact term is required to keep the full amplitude gauge invariant, and can be written as

\begin{eqnarray}
{\cal M}_c & = & i \frac{e g_{K \Lambda N} g_{\phi K
K}}{m_N+m_{\Lambda}} F_K(q_{K^-}^2,m_{K^-}) F_K(q_{K^+}^2,m_{K^+})
\nonumber \\
&& \times \bar{u} (p_5,s_5) \gamma_5 \Slash {q}_{K^+} u(p_2,s_2)
G_{K}(q_{K^+})  \varepsilon^*_{\mu}(p_3,s_3) \nonumber \\
&& \times [g^{\mu\nu} - \frac{(p_4+q_{K^+})^{\mu}p_3^{\nu}}{p_1
\cdot p_3}]\varepsilon_{\nu}(p_1,s_1).
\end{eqnarray}

Then the calculations of the differential and total cross sections
for $\gamma p \to \phi K^+ \Lambda$ reaction are straightforward,
\begin{eqnarray}
&& d\sigma (\gamma p \to \phi K^+ \Lambda) = \frac{1}{8 E_{\gamma}}
\sum_{s_i} |{\cal M}|^2  \times \nonumber\\ &&\frac{d^{3} p_3}{2
E_3} \frac{d^{3} p_4}{2 E_4} \frac{m_\Lambda d^{3} p_5}{E_5}
 \delta^4(p_1+p_2-p_3-p_4-p_5),
\label{eqcs}
\end{eqnarray}
where $E_3$, $E_4$ and $E_5$ are the energy of the $\phi$, $K^+$,
and $\Lambda$, respectively, and $E_{\gamma}$ is the photon energy
at the laboratory frame. Since the relative phase between various
amplitudes is not known, the interference terms between these parts
are ignored in the present calculation.~\footnote{In effective
Lagrangian approaches, the relative phases between different
amplitudes are not fixed. We should generally introduce a relative
phase between different amplitudes as a free parameter. However, we
do not have experimental information for the $\gamma p \to \phi K^+
\Lambda$ reaction, and we will see in the following that the
magnitudes of the contributions from different processes in the
energy region what we considered are much different, hence the
effect of the interference term should be small, and we ignore them
in the present work.}

\section{Numerical results and discussions}

With the formalism and ingredients given above, the total cross
section versus the beam energy $E_{\gamma}$ for the $\gamma p \to
\phi K^+ \Lambda$ reaction is calculated by using a Monte Carlo
multi-particle phase space integration program.~\footnote{In our
calculations, we take $g^2_{N^*(1535)K\Lambda} = 3.52$. Because the
value of $R$ varies in a wide range as mentioned in the
Introduction, we take $R = 1$ here for simplicity. $R = 1$ leads to
$g^2_{K\Lambda N^*(1535)} = R^2 g^2_{\eta N N^*(1535)} = 3.52$ based
on Table~\ref{tab1}.}

\begin{figure}[htbp]
\begin{center}
\includegraphics[scale=1.0]{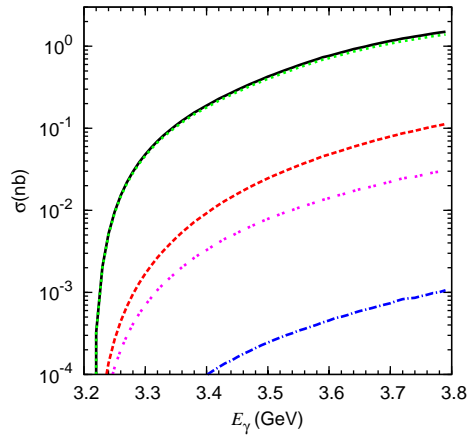}
\caption{(Color online) Total cross sections for the $\gamma p \to
\phi K^+ \Lambda$ reaction as a function of the beam energy
$E_{\gamma}$. The red-dashed, green-dotted, and blue-dashed-dotted
lines stand for contributions from $\pi$, $\eta$, $K$ and contact
term, respectively. Their sum is shown by the solid line. The
pink-dotted-dotted curve is obtained with $\Lambda_K =1.0$ GeV.}
\label{Fig:tcsmeson}
\end{center}
\end{figure}

The roles of various meson exchange processes in describing the
total cross section are shown in Fig.~\ref{Fig:tcsmeson}, where one
can see that the $\eta$ meson exchange plays a predominant role,
while the contributions from the $\pi$ exchange, $K$ exchange, and
contact term are small. This behavior does not vary much with $R
\equiv | g_{K \Lambda N^*(1535)}/g_{\eta N N^*(1535)} |$. The
overwhelming $\eta$ exchange contribution, compared with that from
the $\pi$ exchange, can be easily understood since the value of the
coupling constant $g^2_{\phi \gamma \eta}$ is 26 times lager than
the one of $g^2_{\phi \gamma \pi}$. Hence, this reaction provides a
good platform for studying the nucleon resonances that couple
strongly to the $\eta N$ and $K \Lambda$ channels.

It is worthy mentioning that the contribution from the $K$ exchange
and contact term is small, but it increases rapidly and depends much
on the value of the cut off parameter $\Lambda_K$. To see how much
it depends on the cut off parameter, we also show in
Fig.~\ref{Fig:tcsmeson} the theoretical results with $\Lambda_K =
1.0$ GeV for comparison.

\begin{figure}[htbp]
\begin{center}
\includegraphics[scale=1.0]{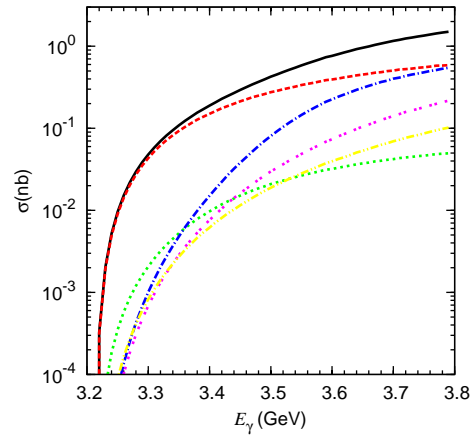}
\caption{(Color online) Total cross sections vs the beam energy
$E_{\gamma}$ for the $\gamma p \to \phi K^+ \Lambda$ reaction. The
red-dashed, green-dotted, blue-dashed-dotted, and pink-dotted-dotted
and yellow-dotted-dotted-dashed lines stand for contributions from
$N^*(1535)$, $N^*(1650)$, $N^*(1710)$, $N^*(1720)$ and background,
respectively. Their sum is shown by the solid line.}
\label{Fig:tcsnstar}
\end{center}
\end{figure}

The relative importance of the contributions of each intermediate
resonance to the $\gamma p \to \phi K^+ \Lambda$ reaction is
demonstrated in Fig.~\ref{Fig:tcsnstar}, where the contributions of
$N^*(1535)$, $N^*(1650)$, $N^*(1710)$, $N^*(1720)$, and background
are shown by red-dashed, green-dotted, blue-dashed-dotted, and
pink-dotted-dotted and yellow-dotted-dotted-dashed curves,
respectively. Their total contribution is depicted by the solid
line. It is clear that the $N^*(1535)$ resonance gives the dominant
contribution from the reaction threshold to $E_{\gamma}$ around
$3.4$ GeV, while the other resonances and background give the minor
contribution. When the beam energy $E_{\gamma}$ is above $3.4$ GeV,
the contributions from other processes are also important.

The contribution from the $N^*(1535)$ resonance be proportional to
$R^2$. If $R$ varies in the range of about $0.5 \sim 2.6$ as
mentioned in Sec.~\ref{introduction}, from Fig.~\ref{Fig:tcsnstar}
one can see that near the threshold of the $\gamma p \to \phi K^+
\Lambda$ reaction, the contribution from the $N^(1535)$ resonance
remains dominant.

When the $\eta$ exchange is dominant as shown in
Fig.~\ref{Fig:tcsmeson}, these facts, i.e., the $N^*(1535)$
resonance couples to $\eta N$ and $K \Lambda$ channels in the $S$
wave, the $N^*(1710)$ and $N^*(1720)$ resonances couple to $\eta N$
and $K\Lambda$ channels in $P$ wave, and the $N^*(1650)$ couples
weaker to $\eta N$ channel, could account for the fact that the
contribution from the $N^*(1535)$ resonance is dominant, while the
contributions from the $N^*(1650)$, $N^*(1710)$, and $N^*(1720)$ are
suppressed at the beam energy closed to the reaction threshold.

\begin{figure}[htbp]
\begin{center}
\includegraphics[scale=0.45]{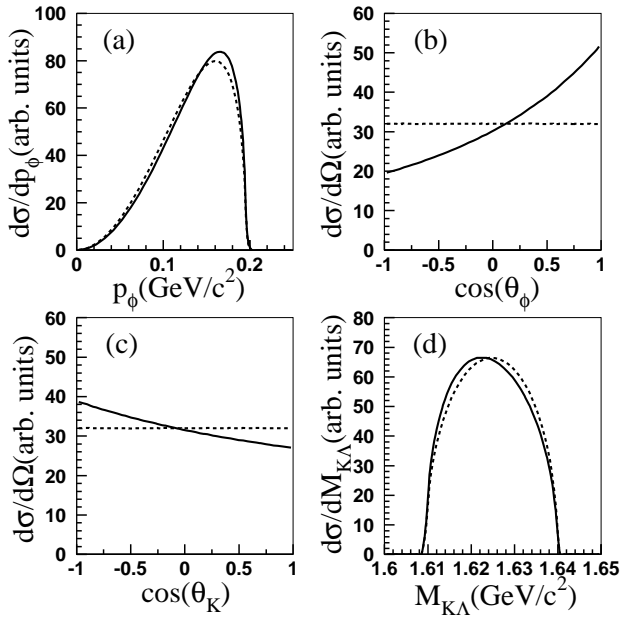}
\vspace{-0.2cm} \caption{Differential distributions for the $\gamma
p \to \phi K^+ \Lambda$ reaction at the beam energy $E_{\gamma}$ =
3.3 GeV. The solid curves stand for our theoretical predictions
while the dashed lines represent the pure phase space
distributions.} \label{phik33}
\end{center}
\end{figure}

\begin{figure}[htbp]
\begin{center}
\includegraphics[scale=0.45]{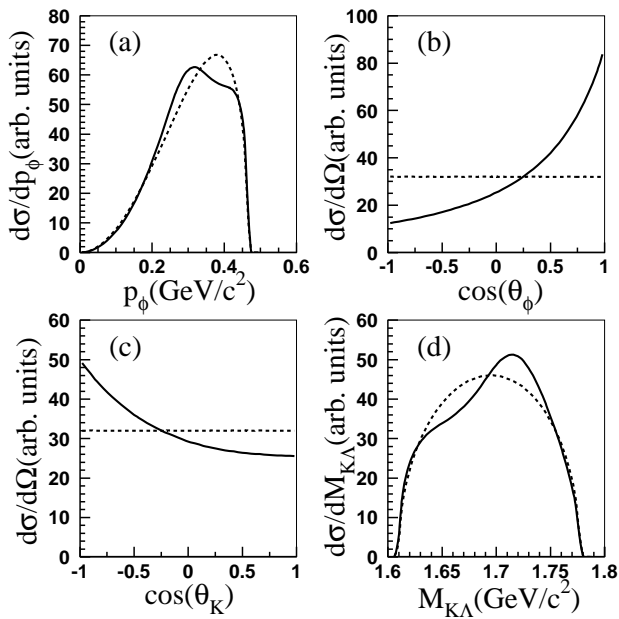}
\vspace{-0.2cm} \caption{As in Fig.~\ref{phik33} but at the beam energy $E_{\gamma}$ = 3.7 GeV.} \label{phik37}
\end{center}
\end{figure}

In addition to the total cross sections, the differential
distributions for $\gamma p \to \phi K^+ \Lambda$ reaction are also
calculated. The corresponding momentum of $\phi$ meson, the angular
distributions of $\phi$ and $K^+$ meson, and the $K \Lambda$
invariant mass spectrum at beam energy $E_{\gamma} = $ 3.3, and 3.7
GeV are shown in Figs.~\ref{phik33} and \ref{phik37}, respectively.
The dashed lines are pure phase space distributions and the solid
lines stand for our theoretical results.

From Figs.~\ref{phik33} and \ref{phik37}, one can see the difference
between the momentum distribution of $\phi$ meson and phase space
distribution is slight at $E_{\gamma} = $ 3.3 GeV but apparent at
$E_{\gamma} = $ 3.7 GeV. Also, the special features of the angular
distributions of $\phi$ and $K^+$ mesons, i.e., the forward
contribution for $\phi$ meson and the backward contribution for
$K^+$ meson, exist at both beam energies. For the $K \Lambda$
invariant mass spectrum, the enhancement near threshold is from the
contribution of $N^*(1535)$ resonance and the other bump appears at
$E_{\gamma} = $ 3.7 GeV is produced by the $N^*(1710)$ resonance.

\section{Summary}

In this work, we have studied the $\gamma p \to \phi K^+ \Lambda$
reaction near threshold within an effective Lagrangian approach. In
addition to the background contributions from the $s$-channel
nucleon pole, $K$ exchange, and contact term, the intermediate
nucleon resonances due to the $\pi$ and $\eta$ meson exchanges are
also investigated. The total and differential cross sections are
predicted. Our results show that the contribution from $N^*(1535)$
resonance due to $\eta$ exchange play the dominant role near
threshold, while other resonances and background contributions are
small and can be ignored. Thus, this reaction provides a good chance
to study the coupling of $K \Lambda N^*(1535)$ interaction. It is
also found that the $\phi$ meson has the forward angular
distribution, while $K^+$ meson gives the backward contribution.
These specific features of angular distributions, together with the
total cross section which is in the magnitude of $0.2$ nb at photon
energy $E_{\gamma} = 3.3 \sim 3.4$ GeV can be tested future
experiments. An experiment with a precision about $0.1$ nb will be
enough to check our model. The future experiment in the JLab 12 GeV
upgrade with large luminosity promise to reach such a requirement.

\section*{Acknowledgments}

We would like to thank Xu Cao for useful discussions. This work is
partly supported by the Ministry of Science and Technology of China
(2014CB845406), the National Natural Science Foundation of China
under grants: 11105126, 11475227, 11375024 and 11175220. We
acknowledge the one Hundred Person Project of Chinese Academy of
Science (Y101020BR0). The Project is Sponsored by the Scientific
Research Foundation for the Returned Overseas Chinese Scholars,
State Education Ministry.

\bibliographystyle{unsrt}

\end{document}